%***Plain TeX file
%
%\AGLTR.TEX
%\nopagenumbers
\magnification 1200
\hsize 145 true mm
\vsize 226 true mm
\magnification=\magstep1
\baselineskip=18truept
\parindent=20truept
\parskip=0truept
\rightline{MIT-CTP-2498, NTUTH-96-01, \& hep-ph/9601219} 
\medskip
\centerline{\bf QCD Sum Rules and Chiral Symmetry Breaking}
\medskip
\centerline{W-Y. P. Hwang}
\smallskip
\parindent=0truept
\centerline{Department of Physics, National Taiwan University}

\centerline{Taipei, Taiwan 10764, R.O.C.}

\centerline{and}

\centerline{Center for Theoretical Physics}

\centerline{Laboratory for Nuclear Science and Department of Physics}

\centerline{Massachusetts Institute of Technology}

\centerline{Cambridge, Massachusetts 02139}
\smallskip
\centerline{\it (revised: 25 October 1996)}
 
\bigskip
\centerline{\bf Abstract}
\medskip
\parindent=20truept
\baselineskip=24truept

We investigate, in the context of QCD sum rules, the role of Goldstone bosons
arising from chiral symmetry breaking, especially on the need of matching 
the world of hadrons to that of some effective chiral quark theory.
We describe how such matching yields predictions on induced condensates 
which reproduce the observed value of the strong $\pi NN$ coupling 
constant. Our result indicate that the observed large $\pi NN$ coupling comes
primarily from the nonperturbative induced-condensate effect. On the other 
hand, the prediction on the parity-violating weak $\pi NN$ coupling,
$ f_{\pi NN} \sim (3.0\pm 0.5) \times 10^{-7}$ is
in good agreement with the value ($\sim 2 \times 10^{-7}$) obtained by 
Adelberger and Haxton from an
overall fit to the existing data, but may appear to be slightly too
large (but not yet of statistical significance) as compared to the most 
recent null experiment. Altogether, we wish to argue that the proposal of 
matching onto a chirally broken effective theory should 
enable us to treat, in a quantitative and consistent manner, the various 
nonleptonic strong and weak processes involving Goldstone bosons.

\bigskip
\parindent=0truept
PACS Indices: 12.39.Fe; 12.38.Lg; 12.38.Aw.
 
\vfill\eject
 
\parindent=20truept
\centerline{\bf I. Introduction}
\medskip

The problem of strong interaction physics has been with us for more than half 
a century, despite the fact that the very nature of the problem has been 
changing 
with the so-called ``underlying theory'' which nowadays is taken
universally to be quantum chromodynamics or QCD. Although the asymptotically 
free nature of QCD allows us to test the candidate theory at high energies, 
the nonperturbative feature dominates at low energies so that it remains 
almost impossible to solve problems related to hadrons or nuclei, except 
perhaps through lattice simulations but it is unlikely that, through the 
present-day computation power which is still not quite adequate, 
simulations would give us most of what we wish to know. As a result,   
it would be useful to evaluate the merits of using the method of QCD sum rules
[1] to bridge between the description of the hadron properties and what we may
expect from QCD. 

The ground state, or the vacuum, of QCD is known to be nontrivial, in the 
sense that there are non-zero condensates, including gluon condensates, 
quark condensates, and perhaps infinitely many higher-order condensates. In 
such a theory, propagators, i.e. causal Green's functions, such as the 
quark propagator
$$ iS_{ij}^{ab}(x) \equiv <0\mid T(q_i^a(x) {\bar q}_j^b(0))\mid 0>, 
\eqno(1)$$
carry all the difficulties inherent in the theory. Higher-order condensates,
such as a four-quark condensate,
$$ <0\mid T({\bar\psi}(z) \gamma_\mu \psi (z) q_i^a(x) {\bar q}_j^b(0))\mid 0>,
$$
with $\psi(z)$ also labeling a quark field, represent an infinite series of
unknowns unless some useful ways for reduction can be obtained. As the vacuum,
$\mid 0>$, is highly nontrivial, there is little reason to expect that Wick's
theorem (of factorization), as obtained for free quantum field theories, is 
still of validity. Thus, we must look for alternative routes in order to 
obtain useful results or relations.

Accordingly, it might be useful, before embarking on the subject of 
chiral symmetry breaking in relation to QCD sum rules, to detour slightly 
by considering the feasibility of working directly with the various matrix
elements (rather than the operators). To this end, we note that, 
at the classical level, we may start with the QCD lagrangian [2],
$$\eqalignno{
  {\cal L}_{QCD} = & {1\over 4} G_{\mu\nu}^a G^{a\mu\nu} - {1\over 2 \xi}
                   (\partial_\mu A^{\mu a})^2 &\cr
               & -{\bar \eta}^a \partial_\mu (\partial^\mu \delta^{ac} 
                   + g f^{abc}A^{\mu b}) \eta^c &\cr
               & + {\bar \psi} \{ i \gamma^\mu (\partial_\mu + ig {\lambda^a
                 \over 2} A_\mu^a) - m\} \psi, &(2)}$$ 
with the gauge-fixing and ghost terms (i.e., the second and third terms) 
shown explicitly. We then have the equations for interacting fields,
$$\eqalignno{
&\{i \gamma^\mu (\partial_\mu + ig {\lambda^a\over 2} A_\mu^a) - m \}\psi =0; 
 &(3)\cr
&\partial^\nu G_{\mu\nu}^a - 2 g f^{abc} G^b_{\mu\nu}G^c_\nu 
 +g {\bar \psi} {\lambda^a\over 2}\gamma_\mu \psi = 0, &(4)}$$
which may be re-interpreted as the equations of motion for {\it quantized} 
interacting fields. As the (standard) rule for quantization, the equal-time 
(anti-)commutators among these quantized interating fields are identical to
those among non-interacting quantized fields.  

Allowing the operator $\{i \gamma^\mu \partial_\mu - m \}$ to act on the 
matrix element defined by Eq. (1), we obtain, making use of what we have
just stated about interacting field equations and (canonical) 
quantization rules, 
$$\eqalignno{
\{i \gamma^\mu \partial_\mu - m \}_{ik} iS_{kj}^{ab} (x) 
  = & i \delta^4(x) \delta^{ab} \delta_{ij} 
     +<0\mid T(\{g {\lambda^n\over 2}A_\mu^n\gamma^\mu q(x)\}_i^a 
     {\bar q}_j^b(0))\mid 0>. &\cr &&(5)}$$
This equation can be solved by splitting the propagator into a singular,
perturbative part and a nonperturbative part:
$$iS_{ij}^{ab} (x) = i S_{ij}^{(0)ab}(x) + i {\tilde S}_{ij}^{ab}(x), 
\eqno(6)$$
with 
$$\eqalignno{
  iS_{ij}^{(0)ab}(x) &\equiv \int {d^4p\over (2\pi)^4} e^{-ip\cdot x}   
    iS_{ij}^{(0)ab}(p), &(7a)\cr
  iS_{ij}^{(0)ab}(p) &= \delta^{ab} {i ({\not p} + m)_{ij}\over p^2 -m^2 +i
   \epsilon}. &(7b)}$$
Accordingly, we have
$$\{i \gamma^\mu \partial_\mu - m \}_{ik} i{\tilde S}_{kj}^{ab} (x) 
  =  <0\mid T(\{g {\lambda^n\over 2}A_\mu^n\gamma^\mu q(x)\}_i^a 
     {\bar q}_j^b(0))\mid 0>. \eqno(8)$$
With the fixed-point gauge, 
$$ A_\mu^n(x) = -{1\over 2} G_{\mu\nu}^n x^\nu + \cdot\cdot\cdot, \eqno(9)$$
we may then solve the nonperturbative part $i{\tilde S}_{ij}^{ab}(x)$ as a
power series in $x^\mu$, 
$$\eqalignno{
i{\tilde S}_{ij}^{ab}(x) = & -{1\over 12} \delta^{ab}\delta_{ij} <{\bar q}q>
   + {i\over 48} m (\gamma_\mu x^\mu)_{ij}\delta^{ab} <{\bar q}q> &\cr
    &+{1\over 192} <{\bar q} g \sigma\cdot G q> \delta^{ab} x^2 \delta_{ij}
     +\cdot\cdot\cdot, &(10)}$$
The first term is the integration constant which defines the so-called
``quark condensate'', while the mixed quark-gluon condensate appearing in
the third term arises because of Eqs. (8) and (9). It is obvious that 
the series (10) is a short-distance expansion, which converges
for small enough $x_\mu$. 

A practitioner would recognize immediately that Eq. (10) is just the standard 
quark propagator cited in most papers in QCD sum rules. The approach which we 
suggest here is based upon two key elements, namely, the set of interacting 
field equations {\it plus} the rule of canonical quantization (for interacting 
fields). The equations which we obtain, such as Eq. (5), are much the same as
the set of Schwinger-Dyson equations (for the matrix elements). An important
aspect in our derivation is that the nontriviality of the vacuum $\mid 0>$ is
observed at every step $-$ a central issue in relation to QCD. Although Eq. 
(10) seems to be standard (and natural), our approach makes it intuitive and
relatively straightforward to work out the problem to an arbitrary order as
well as to treat many other (propagator-like) matrix elements such as those
to be used later in the paper.

In light of the fact that the vacuum $\mid 0>$ is nontrivial (a fact which
renders useless many lessons from perturbative quantum field theory), the 
route of starting directly from the interacting field equations, augmented by
the canonical quantization rule, may be pursued
seriously. The method, in which renormalizations need to be defined order by 
order in a consistent manner, may be contrasted with the usual path-integral 
formulation in which Green's functions may be derived, in an elegant manner, 
from the generating functional. The method should be seriously exploited 
mainly because, just like what we have done in the above, it might offer some 
new insights for the problem at hand, i.e. QCD.

As another important exercise, we may split the gluon propagator into the 
singular, perturbative part and the nonperturbative part. For the 
nonperturbative part, our interacting-field-equation-based approach yields
$$\eqalignno{
 & g^2 <0\mid \colon G_{\mu\nu}^n(x) G_{\alpha\beta}^m(0) \colon \mid 0> &\cr
= &\quad {\delta^{nm}\over 96} <g^2G^2> (g_{\mu\alpha}g_{\nu\beta} 
    -g_{\mu\beta}g_{\nu\alpha}) &\cr
  &- {\delta^{nm}\over 192} <g^3 G^3> \{ x^2 (g_{\mu\alpha}g_{\nu\beta} -
  g_{\mu\beta}g_{\nu\alpha}) - g_{\mu\alpha}x_\nu x_\beta + g_{\mu\beta}x_\nu
    x_\alpha -g_{\nu\beta}x_\mu x_\alpha +g_{\nu\alpha} x_\mu x_\beta \} &\cr
  & + O(x^4), &(11)}$$
with 
$$\eqalignno{
 <g^2 G^2 > & \equiv <0\mid \colon g^2 G_{\mu\nu}^n (0) G^{n\mu\nu}(0)
    \colon \mid 0>, & (12a)\cr
 <g^3 G^3> & \equiv <0\mid \colon g^3 f^{abc}g^{\mu\nu} G_{\mu\alpha}^a(0)
    G^{b\alpha\beta}(0) G_{\beta\nu}^c(0) \colon \mid 0>.&(12b)}$$
Again, the first term in Eq. (11) is the integration constant for the 
differential equation satisfied by the gluon propagator. Note that inclusion
of the gluon condensate $<g^2 G^2 >$ in Eq. (11) is still standard but 
the triple gluon condensate $<g^3 G^3 >$ is a new entry required by the 
interacting field equation (4). Our approach indicates when condensates of 
entirely new types should be introduced as we try to perform operator-product 
expansions to higher dimensions.

The method of QCD sum rules may now be regarded as the various approaches
in which we try to exploit the roles played by the quark and gluon condensates
which enter Eqs. (10) and (11) for problems involving hadrons. To this end,
we wish to set up a quantum field theory in which the vacuum is nontrivial, 
while leaving some condensates as parameters to be determined {\it via} first
principle, such as from lattice simulations of QCD. The aim is such that, 
once the fundamental condensate parameters are given, one can make 
predictions on the 
various physical quantities associated with hadrons. Of course, the method of
QCD sum rules in its present broad context has its orgin in [1].

\bigskip
\bigskip
\centerline{\bf II. Induced Condensates and the Effective Chiral Quark Theory}
\medskip
\medskip

The quark condensate which enters the various QCD sum rules, such as 
the Belyaev-Ioffe nucleon mass sum rule [3] and the sum rules [4] for 
the nucleon axial couplings, is perhaps the best known parameter among
all condensates:
$$ a \equiv -(2\pi)^2 <{\bar q} q> \simeq 0.546\, GeV^3,\eqno (13)$$
which is the order parameter in the chirally broken phase of QCD. It is
related to the pion decay constant $f_\pi$ with $4\pi f_\pi$ being the
scale governing the convergence property of the expansion in chiral 
perturbation theory. One point which is worth mentioning at this juncture
is that, as we vary $Q^2$ in probing the hadron
substructure, the nonzero condensate as listed in Eqs. (13) does not 
disappear suddenly although the effects may become negligibly small for
sufficiently large $Q^2$ $-$ unlike the situation when one varies the 
temperature $T$ and sees some phase transition at some critical 
temperature $T_c$. As long as we submit to the picture where the
condensate (13) differs from zero, it is important to recognize that we are
working with a condensated phase of QCD $-$ which already differs from 
the naive QCD (without condensation). It is also natural to ask what other
effects such condensation (phase transition) brings to light. For instance,
it is believed that the spontaneous symmetry breaking leading to sizable 
condensates also gives rise to the dynamically-generated or
constituent mass to a quark. Such 
spontaneous symmetry breaking refers to the breaking of $SU(N_f)_L \times 
SU(N_f)_R$ chiral symmetry for QCD of $N_f$ flavors of almost massless quarks 
into the $SU(N_f)_V$ symmetry which, according to Goldstone theorem, must give
rise to $N_f^2-1$ almost massless Goldstone bosons, pseudoscalar bosons in the
present case (with pions and other pseudoscalar mesons as the natural 
candidate). If we restrict ourselves to the world of up and down quarks (as 
we shall do in the rest of this paper), there is little reason to reject the
role played by Goldstone pions while accepting quite large empirical values 
for the various condensates.

In light of the above consideration, we are led to investigate, in the context 
of QCD sum rules, the role of Goldstone bosons arising from chiral symmetry 
breaking, especially on the need of matching the world of hadrons to that of 
some effective chiral quark theory. We immediately recognize that such
proposal has many raminifications for the method of QCD sum rules.
In any event, the 
proposal should first be checked out to leading order in the 
Goldstone fieldss $-$ such as very small Goldstone pion fields. A natural 
problem for such investigation is the determination of the strong $\pi NN$ 
coupling, for which we shall show that the proposal indeed yields predictions 
on induced condensates which are just needed to reproduce the observed value 
and that such large $\pi NN$ coupling comes primarily from the nonperturbative 
induced-condensate effect. A closely related question is the parity-violating
(p.v.) weak $\pi NN$ coupling which is considerably more complicated since it
involves calculation (and regularization) of three-loop diagrams. In any
event, it is of importance to use the determination of the $\pi NN$ 
couplings as the first testing ground for the proposal. This is what we 
wish to do in the present paper, while having to put off investigations of 
some other serious issues for the future.

What should we use while still insisting on using quarks and gluons as the 
effective degrees of freedom (DOF's)? A plausible choice is to accept the 
chiral lagrangian of Weinberg [5] and Georgi [6] as the candidate effective
theory, in which Goldstone bosons arising from chiral symmetry breaking 
couple to quarks (and indirectly to gluons) but such couplings should vanish 
in the long-wavelength limit. The choice may not be unique as 
far as identification of Goldstone pions (with physical pions) is concerned, 
but many different choices may turn out to be equivalent (as some people 
believe to be the case).
 
On the other hand, it is often said that the long-distance realization, or 
the low-energy effective theory, of QCD is chiral perturbation theory 
($\chi$PT) [7], in which
Goldstone bosons are the only effective DOF's. As we go up in the energy 
scale, we must eventually end up in having to deal directly with QCD itself.
Thus, it is also natural to anticipate that, at some intermediate energy
scale, the two languages, namely $\chi$PT and QCD, can be matched onto
each other or, in somewhat loose terms, the DOF's of the two theories couple
in a way dictated by symmetries, yielding again the effective chiral quark
theory [5,6]. The resultant theory is unique if symmetries dictates 
completely the interactions, as suspected to be so in the present case. 

Therefore, there leaves little room for what we may choose for the 
effective chiral quark theory [5,6], onto which we shall try to map the 
world of hadrons. As the primary example of this paper, we consider the
breaking of the $SU(2)_L \times SU(2)_R$ chiral symmetry into the vector 
$SU(2)$ symmetry. The effective chiral quark theory is then specified, to 
dimension six, by [5]

$$\eqalignno{
{\cal L} = &\quad {\bar \psi}\{ i \gamma^\mu \partial_\mu - {1\over 4 f_\pi^2}
\gamma^\mu (\partial_\mu {\vec \pi}) \times {\vec \pi} \cdot {\vec \tau} \}
\psi &\cr
& -{g_A\over 2 f_\pi} {\bar \psi} \gamma^\mu\gamma_5 {\vec \tau}\psi \cdot
\partial_\mu{\vec \pi} - m {\bar \psi} \psi + {1\over 2} (1- {{\vec \pi}^2
\over 4 f_\pi^2})\partial^\mu {\vec\pi}\cdot\partial_\mu{\vec \pi}&\cr
& - {\mu^2\over 2}(1- {{\vec \pi}^2 \over 4 f_\pi^2}) {\vec \pi}^2 
- {\bar\psi} \Gamma_\alpha \psi {\bar\psi}\Gamma^\alpha \psi, &(14)}$$
to which the gluon DOF's should be added as in Eq. (2), but with considerably
weaker coupling constant.

The method of QCD sum rules starts with a choice of some correlation function
which may be evaluated at the quark level on the one hand (the so-called 
``left-hand side'' or LHS) while may also be interpreted at the hadron
level (the so-called ``right-hand side'' or RHS). To study the nucleon
properties, for instance, we may choose
$$\Pi(p) = i\int d^4x e^{ip\cdot x}<0|T[\eta_p(x)\bar\eta_p(0)]|0>\;,
\eqno(15a)$$
with the usual choice [8] of the current $\eta(x)$
$$\eta_p (x)  = \epsilon^{abc}[u^{aT}(x)C\gamma_\mu u^b(x)]\gamma^5
\gamma^\mu d^c(x). \eqno(15b)$$
Here $C$ is the charge conjugation operator and $<0\mid \eta (0)\mid N(p)> 
= \lambda_N u_N(p)$ with $\lambda_N$
the amplitude for overlap and $u_N(p)$ the nucleon Dirac spinor (normalized
such that ${\bar u}_N(p)u_N(p) = 2m_N$). And we may work out both the LHS and
RHS sides and obtain the now well-known QCD sum rules [3, 4].

This is ``duality'' in the sense that some quantity assumes suitable meaning
both at the quark level and at the hadron level. As elucidated in detail so 
far, we wish to investigate the proposal that, in such 
duality picture, the language to be used at the quark level is the 
effective chiral quark theory, rather than QCD in the chirally symmetric 
phase (in which Goldstone bosons do 
not enter as relevant DOF's). Since chiral symmetry breaking has 
already taken place with the order parameter as given by Eq. (13),
it appears quite natural that we should be trying to match the world of 
hadrons to a world in which we could talk about not only quarks and 
gluons but also sizable basic condensates. 

To proceed further, we note [9] that, on an empirical ground, the quark 
propagator is modified in the presence of sufficiently small pion fields:
$$\eqalignno{
i\delta S^{ab}(x)  = & -{i\vec\tau\cdot\vec\pi\over 4\pi^2x^2}\;g_{\pi q}
\gamma^5\delta^{ab} &\cr
 & +{i\over 24}\;\vec\tau\cdot\vec\pi\;g_{\pi q}\chi_\pi<\bar
qq>\delta^{ab}\gamma^5\;, &\cr
 & - {i\over 3\cdot 2^7}\;m_{0}^{\pi}<\bar qq>g_{\pi q}
  \vec\tau\cdot\vec\pi x^2 \gamma^5, &(16)}$$
with
$$\eqalignno{
<\bar qi\tau_j\gamma_5q>_\pi &\equiv
\chi_\pi g_{\pi q}\pi_j<\bar qq> &(17a)\cr 
<\bar q\;i \gamma_5 \tau_j \sigma\cdot G q>_\pi &\equiv
m_{0}^{\pi}g_{\pi q}\pi_j <\bar qq>. &(17b)}$$
Here $g_{\pi q}$ is the pion-quark coupling, which has been introduced on a
purely empirical ground. The
susceptibility $\chi_\pi$ enters in the evaluation of both strong and weak
pion-nucleon coupling constants, while $m^\pi_0$ enters only for the weak one.

We may begin by working in the limit of massless Goldstone pions 
($\mu^2 = 0$). 
In this limit, all the terms involving Goldstone bosons must contain the
derivative $\partial_\mu {\vec \pi}$, as shown by the lagrangian of Eq. (14). 
Denote these terms altogether by $\delta {\cal L}$. For the primary purpose of 
the present paper, it is sufficent to treat such pion fields as some small
classical fields. Thus, we obtain the modification to the quark propagator as 
defined by Eq. (1).

$$\eqalignno{
i\delta S_{ij}^{ab}(x) &= <0\mid T(i\int d^4 z \delta {\cal L}(z) q_i^a(x) 
{\bar q}_j^b(0))\mid 0> &\cr
& = -i \int d^4z\, \partial^\mu {\vec \pi}(z)\cdot <0\mid T({\vec \alpha}_\mu(z)
q_i^a(x) {\bar q}_j^b(0))\mid 0>&\cr
& = i {g_A\over 2 f_\pi}\int d^4z\, {\vec \pi}(z)\cdot \partial^\mu <0\mid 
T({\vec A}_\mu(z) q_i^a(x) {\bar q}_j^b(0))\mid 0>,&(18)\cr}$$
with ${\vec A}_\mu(x)$ the isovector axial vector current as given by
$${\vec A}_\mu = {\bar \psi}\gamma^\mu \gamma_5 {\vec \tau}\psi 
   -{2f_\pi\over g_A} (1- {{\vec \pi}^2\over 4 f_\pi^2}) \partial^\mu {\vec \pi}
   -{1\over 2 g_A f_\pi} {\bar\psi}\gamma^\mu {\vec \tau}\times {\vec \pi}\psi.
  \eqno(19)$$
Note that the second equality in Eq. (18) follows since we shall treat pions 
fields $\vec \pi(x)$ as some small classical field. 
It is straightforward to show that PCAC holds:
$$ \partial_\mu {\vec A}^\mu = {2 f_\pi\over g_A} \mu^2 {\vec \pi} + O({1\over
   f_\pi}).\eqno(20)$$
A little algebra yields, in the limit with $\mu^2=0$,
$$\eqalignno{
i\delta S_{ij}^{ab}(x) =& -i {g_A\over 2 f_\pi} {\vec \pi}(x)\cdot <0\mid 
T(\{\gamma_5 {\vec \tau} q(x)\}_i^a {\bar q}_j^b(0))\mid 0> &\cr
& -i {g_A\over 2 f_\pi} {\vec \pi}(0)\cdot <0\mid 
T( q_i^a(x) \{{\bar q}(0)\gamma_5 {\vec \tau})\}_j^b\mid 0>. & (21)\cr}$$
This is in fact a notable case in which the expectation value of the product
of some four-quark operators can be reduced to an expression involving only
two-quark operators $-$ via the application of PCAC.

The two propagators appearing in Eq. (21) can be solved in a way 
introduced in the previous section for 
the propagator $i S_{ij}^{ab}(x)$. Namely, we use Eq. (3) and equal-time 
anti-commutators and obtain
$$\eqalignno{
 &\{i \gamma^\mu \partial_\mu + m \}_{ik} 
<0\mid T(\{\gamma_5 {\vec \tau} q(x)\}_k^a {\bar q}_j^b(0))\mid 0> &\cr
  = & - i \delta^4(x) \delta^{ab} \gamma^5{\vec\tau} 
     -<0\mid T(\{\gamma^5{\vec\tau} g {\lambda^n\over 2}A_\mu^n\gamma^\mu 
     q(x)\}_i^a {\bar q}_j^b(0))\mid 0>, &(22a)\cr
& \{i \gamma^\mu \partial_\mu - m \}_{ik}
<0\mid T( q_k^a(x) \{{\bar q}(0)\gamma_5 {\vec \tau})\}_j^b\mid 0> &\cr
  = & i \delta^4(x) \delta^{ab} \gamma^5 {\vec\tau}
     +<0\mid T(\{g {\lambda^n\over 2}A_\mu^n\gamma^\mu q(x)\}_i^a 
     \{{\bar q}(0)\gamma^5{\vec\tau}\}_j^b(0))\mid 0>. &(22b)}$$
The solutions to Eqs. (22a) and (22b) are
$$\eqalignno{
&<0\mid T(\{\gamma_5 {\vec \tau} q(x)\}_i^a {\bar q}_j^b(0))\mid 0> &\cr
  = & \int {d^4p\over (2\pi)^4} e^{-ip\cdot x}   
\delta^{ab} {i [(-{\not p} + m)\gamma^5{\vec\tau}]_{ij;mn}\over p^2 -m^2 +i
   \epsilon} 
 -{1\over 12} \delta^{ab}\gamma^5{\vec\tau} <{\bar q}q>
   - {i\over 48} m \gamma_\mu x^\mu\gamma^5{\vec\tau}\delta^{ab} 
   <{\bar q}q> &\cr
    &+{1\over 192} <{\bar q} g \sigma\cdot G q> \delta^{ab}\gamma^5{\vec\tau}
     x^2 +\cdot\cdot\cdot, &(23a)\cr
&<0\mid T( q_i^a(x) \{{\bar q}(0)\gamma_5 {\vec \tau})\}_j^b\mid 0> &\cr
  = & \int {d^4p\over (2\pi)^4} e^{-ip\cdot x}   
\delta^{ab} {i [({\not p} + m)\gamma^5{\vec\tau}]_{ij;mn}\over p^2 -m^2 +i
   \epsilon}
 -{1\over 12} \delta^{ab}\gamma^5{\vec\tau} <{\bar q}q>
   + {i\over 48} m \gamma_\mu x^\mu\gamma^5{\vec\tau}\delta^{ab} 
   <{\bar q}q> &\cr
    &+{1\over 192} <{\bar q} g \sigma\cdot G q> \delta^{ab}\gamma^5{\vec\tau}
     x^2 +\cdot\cdot\cdot, &(23b)\cr}$$
We caution that the quark mass $m$ in these equations is the one appearing
in the effective chiral quark theory, Eq. (14), rather than that in the 
bare QCD, Eq. (2) [the chirally symmetric phase of QCD].

Substituting Eqs. (23a) and (23b) back into Eq. (21) and treating ${\vec\pi}$ 
as a small constant field, we obtain

\vfill\eject

$$\eqalignno{
& i\delta S^{ab}(x) &\cr
=& +{g_A m\over f_\pi} {\vec\tau}\cdot {\vec\pi} \int {d^4p\over (2\pi)^4}
   e^{-ip\cdot x} {\delta^{ab}\gamma_5\over p^2 -m^2 +i\epsilon}&\cr
 & + i{g_A\over f_\pi} {\vec \tau}\cdot{\vec\pi} \{ {1\over 12} <{\bar q}q>
   \delta^{ab}\gamma_5 -{1\over 192} <{\bar q} g \sigma\cdot G q>\delta^{ab}
   x^2 \gamma_5\}.& (24)}$$
In the effective chiral quark theory [6], we use $m \approx 0.35 \, GeV$. As 
we usually take $p^2 \sim 1\, GeV^2$, we may neglect $m^2$ as compared 
to $p^2$ in the first term of Eq. (24) so that the expression (24) reduces
to Eq. (16) with
$$\eqalignno{
g_{\pi q} &= -{g_A m\over f_\pi}, &(25a)\cr
\chi_\pi g_{\pi q} &= {2 g_A\over f_\pi}, &(25b)\cr
m_0^\pi <{\bar q}q> g_{\pi q} &= {2g_A\over f_\pi} 
<{\bar q} g \sigma\cdot G q>.  &(25c)}$$
Accordingly, the pion-quark coupling $g_{\pi q}$ and the induced 
condensates as characterized by the susceptibilities $\chi_\pi$ and
$m_0^\pi$ are completely determined when the connection with the effective
chiral quark theory is made.
Numerically, we may use [6]
$$f_\pi = 93\, MeV, \qquad g_A=0.7524, \eqno(26a)$$
so that
$$ g_{\pi q} \sim -2.83, \qquad \chi_\pi a = - {2a\over m} 
\approx -3.14 \, GeV^2.\eqno(26b)$$
Note that we also have $m_0^\pi = {2m_0^2\over m}$ with $<{\bar q} g 
\sigma \cdot G q> \equiv -m_0^2 <{\bar q}q>$ but we shall discuss the
parameter $m_0^2$ later in the paper. We re-iterate again that, since 
the whole derivation is made by making connection with an effective 
chiral quark theory, it is natural to adopt the constituent quark mass 
in Eqs. (26). 

\vfill\eject

\centerline{\bf III. Pion-Nucleon Couplings}
\medskip
\medskip
To investigate the proposal that the matching is done between the hadron 
world and the effective chiral quark theory [5,6], we wish to consider
the evaluation of both the strong and weak $\pi NN$ coupling constants. Such
evaluation, by itself being of fundamental importance, can readily be 
generalized to treat a large number of physical processes involving Goldstone 
bosons, such as nonleptonic decays of hyperons or heavy mesons ($\Lambda \to 
N\pi$, $D\to K\pi$, etc.). 

To begin, it is useful to recall the Belyaev-Ioffe nucleon mass sum 
rule, taking into account the quark mass terms [3, 10], 
$$\eqalignno{
&{1\over 8}M_B^6 L^{-4/9}E_2 + {1\over 32} <g_c^2 G^2> M_B^2 L^{-4/9}E_0
+{1\over 6} a^2 L^{4/9} - {a^2 m_0^2\over 24 M_B^2} L^{-2/27}&\cr 
-& {maM_B^2\over 4} L^{-4/9} E_0 -{mam_0^2\over 8} L^{-26/27}
= \beta_N^2 e^{-M^2/M^2_B},&(27)\cr}$$
with $\beta_N^2 \equiv {1\over 4} (2\pi)^4 \mid \lambda_N\mid^2$. 
The factors containing $L$, $L = 0.621\ln(10M_B)$, give the evolution in 
$Q^2$ arising from the anamolous dimensions, and the $E_i(M^2_B)$ functions 
take into account excited states to ensure the proper large-$M_B^2$ behavior 
[3, 4].

By analyzing the observed mass splittings in heavy quarkonia, Narison [11]
extracted an updated average value on the gluon condensate:
$$<g_c^2 G^2> = (8.9 \pm 1.1) GeV^2, \eqno (28)$$
which, albeit somewhat larger than the early value obtained by SVZ [1], 
appears to be more appropriate for future QCD sum rule analyses. It is 
unlikely that the gluon condensate suitable for light quark systems would
be smaller than this value (as extracted primarily from charmonium systems).
On the other hand, the quoted value on the quark condensate (as an order
parameter of the phase transition), namely Eq. (13), may have an 
uncertainty as large as 20 \%, should we use the instanton picture of the
QCD vacuum [12] as a guide. Making use of the Ward-Takahashi identities
associated with a specific nonlinear chiral transformation, Jacquot and 
Richert [13] obtain
$$m_{dyn}(\mu) <{\bar q}q> = - {1\over 48\pi^2} <g^2(\mu)G^2>,\eqno(29)$$
where $m_{dyn}(\mu)$ is the dynamically generated quark mass at the scale
$\mu$. This relation also follows from a dilute-gas approximation of the
instanton QCD vacuum. Substituting the current values of the condensates 
into this relation, it is seen that the dynamically generated quark mass
may be interpreted as the constituent quark mass. Since the relation (29) 
would fail badly with the current quark mass as the input, this result may
be regarded as one of the reasons why the constituent or dynamically
generated mass might be more appropriate for QCD sum rule studies. 

It is possible to adopt Eq. (29) as a constraint among the quark mass, 
and the quark and gluon condensates. Such constraint leads to the gluon
condensate which is a few times larger than the original SVZ value [1] but
which is pretty much in line with the results from lattice simulations.

As indicated earlier, it is likely that the mixed quark-gluon condensate 
is smaller than the current value $m_0^2 \sim 0.8 \, GeV^2$ especially when
the effective chiral quark theory is used. Lacking any further guideline 
on the matter, we recall the reduction factor in $\alpha_s$ [6] and 
arbitrarily take $m_0^2 \sim 0.4 \, GeV^2$, but caution 
that further justification is needed for coming up with a better value of
$m_0^2$. The convergence property of the mass sum rule, Eq. (27), would be
a suspect for $m_0^2 \approx 0.8 \, GeV^2$ but becomes very stable when
a smaller value is assumed. In a sense we can make use of the nucleon mass
sum rule to constrain the mixed quark-gluon condensate but the sum rule 
becomes so well behaved for sufficiently small $m_0^2$ that the constraint 
loses its efficiency.

The logarithmic derivative of the LHS of Eq. (27) gives the prediction on
the nucleon mass $M$, a prediction which turns out to be quite stable with
respect to the Borel mass $M_B$. We obtain $M = (0.94\pm 0.02)\,GeV$ for
$M_B = (1.10\pm 0.05)\, GeV$. Substituting this value back into Eq. (27),
we obtain $\beta_N^2 = (0.177\pm 0.005) GeV^6$. An additional error of
at least $\pm 0.4 \, GeV$ for the predicted nucleon mass should be 
understood if we consider the uncertainty in the various condensates.

We turn our attention to the QCD sum rule for the strong $\pi NN$ 
coupling [14, 9].

$$\eqalignno{
&{1\over 4} M_B^6 L^{-4/9}E_2 -{\chi_{\pi} a\over 8} M_B^2 L^{2/9}E_1  
- {11\over 96}<g_c^2 G^2> M_B^2 E_0 +{1\over 3} a^2 L^{4/9} &\cr
& + m a m_0^2 L^{-26/27}\{{3\over 8} \ln {M_B^2\over \mu^2} 
+ {49\over 48}\}
  +{1\over 12} m_0^2 a^2 {L^{-2/27}\over M_B^2} &\cr
= & g_{\pi q}^{-1}\{ g_{\pi NN} + B M_B^2\} \beta^2_N\;e^{-M^2/M^2_B}.
&(30) }$$
Note that the term $B\,M_B^2$ is introduced to absorb some effect from
excited states (with less singular pole behavior). 

Numerically, we obtain $g_{\pi NN} = -(14.8 \pm 0.7)$ for $M_B = (1.10 \pm
0.05)\, GeV$, again a fairly stable result with respect to the Borel mass.
An additional error of at least $\pm 2.0$ should be understood in view of  
the uncertainties involved in the various condensates.

We wish to point out that about 50 \% of the contribution in fact comes 
from the $\chi_\pi a$ term, a nonperturbative effect. This result confirms
the long-standing conviction that pion physics associated with nucleons is
non-perturbative and can only be obtained by taking into account the effects
related to the nontrivial vacuum. Experimentally, we have $\mid 
g_{\pi NN} \mid \approx 13.5$, which is in general agreement with the 
sum rule prediction.

It is indeed remarkable that the observed strong $\pi NN$ coupling has been
reproduced in QCD sum rules primarily because of the inclusion of the 
nonperturbative terms $-$ the first term on the LHS is the leading 
perturbative contribution but it is not of much numerical importance. In 
other words, we have essentially shown that the large $\pi NN$ coupling is
nonperturbative in origin. 

Finally, we may also turn our attention to the weak p.v. pion-nucleon 
coupling, which is defined as follows [15, 16, 17].
$$
{\cal L}_{\pi NN}^{p.v.} = {f_{\pi NN}\over \sqrt2} 
\bar\psi({\vec \tau}\times{\vec \pi})_3 \psi. \eqno(31)
$$
Only charged pions can be emitted or absorbed. Here we modify the QCD 
sum rule which we obtained earlier [9] (and which has some pathological 
behavior) by multiplying both sides (LHS and RHS) by
the factor $(p^2 - M^2)$ immediately before the Borel transform. This 
has helped to produce a QCD sum rule which is very well behaved.
$$\eqalignno{
 {G_F\sin^2\theta_W({17\over 3} - \gamma)
\over 96 \pi^2}\; &\bigl[ (4M^{10}_B - M^2 M_B^8) L^{-4/9}E_3 &\cr
& - (2 M_B^8 - {2\over 3} M^2 M_B^6)\chi_\pi\;aL^{-4/9} E_2 
 - (M_B^6 - {1\over 2} M^2 M_B^4) m_0^\pi\;a\;E_1 L^{-4/9}\bigr]&\cr
 = & g_{\pi q}^{-1}\{f_{\pi NN} +B' M_B^2\} \beta^2_N\; 2 M^2\;
      e^{-M^2/M^2_B}.& (32)}$$
This sum rule still suffers from the fact that the contribution from
the gluon condensate diagrams is yet to be included; these diagrams are
much more complicated to evaluate although they are of the same order 
or smaller than uncertainties of our calculation. 

Numerically, we obtain $f_{\pi NN} = (3.04 \pm 0.01) \times 10^{-7}$ for
$M_B =(1.10 \pm 0.05)\, GeV$, a prediction which is even more stable than
the previous two sum rules (on the nucleon mass and the strong pion-nucleon
coupling). The uncertainties in the condensates and in the terms which have
been neglected amount to at least $\pm 0.5 \times 10^{-7}$. About 50 \% 
of the contribution comes from the nonperturbative $\chi_\pi a$ term.  

Our prediction of $ f_{\pi NN} \sim (3.0 \pm 0.5) \times 10^{-7}$ (at 
$M_B \sim 1.0 \, GeV$) is in good agreement with the value ($\sim 2 
\times 10^{-7}$) obtained by Adelberger and Haxton [15] from an
overall fit to the existing data, but may appear to be slightly too
large (but not yet of statistical significance) as compared to the most 
recent experiment [18]. Once again, we see that the non-perturbative physics
dictates in the prediction of the parity-violating weak $\pi NN$ coupling,
a typical case of nonleptonic weak interactions. Our results suggest the 
need to incorporate such non-perturbative effects into any systematic study
of nonleptonic weak interactions. We suspect that many important issues, such
as the $\Delta I = {1\over 2}$ rule, in relation to nonleptonic weak 
interactions cannot be addressed in any meaningful manner without suitable 
incorporation of non-perturbative condensate effects.

\vfill\eject

\centerline{\bf IV. Discussion and Summary}
\medskip
\medskip

In light of the very encouraging results which we have obtained for 
the nucleon mass and the strong
and weak $\pi NN$ couplings, it is tempting to take seriously 
the proposal that the matching in QCD sum rules is done between the hadron 
world and the effective chiral quark theory [5,6]. In particular, 
non-perturbative QCD effects in a large number of physical 
processes involving Goldstone bosons may now be treated on the same footing.

The uncertainties involved in the various condensates pose some problem
in the analysis of QCD sum rules. As a numerical indicator, we note that
an increase of 10 \% in the gluon condensate produces little change in
the predicted nucleon mass $M$ ($-0.4$ \%) and the p.v. pion-nucleon 
coupling $f_{\pi NN}$ ($-1.5$ \%) but reduces the value of the strong
pion-nucleon coupling $\mid g_{\pi NN} \mid$ by 5.7 \%. A reduction of
$m_0^2$ by 20 \% (on the mixed quark-gluon condensate) produces the 
changes of $-2.1$ \%, $-12.6$ \%, and $+15.1$ \% respectively for $M$,
$\mid g_{\pi NN}\mid$, and $f_{\pi NN}$. On the other hand, a reduction of
the quark condensate by 10 \% increases $M$ by 4 \%, decreases 
$\mid g_{\pi NN}\mid$ by 2.2 \%, and reduces $f_{\pi NN}$ by 12.4 \%.
These results enable us to assign some errors to our predictions cited
in the previous section. Nevertheless, some room remains for an entirely 
different set of parameters. For instance, we may remark that a 
relativistic version of the effective chiral quark theory based on the 
Bethe-Salpeter equation (BSE) and Schwinger-Dyson equation (SDE), such as 
[19], could also lead to similar final numerical results. 
Therefore it is premature to take the present numerics as the credence 
towards only the nonrelativistic chiral quark model [6].

There are also many other important issues related to the proposal. For 
example, the adoption of the constituent quark mass could modify the
existing QCD sum rule calculations, perhaps quite severely in some cases. 
Furthermore, pions may also be treated as ``quantized fields'', leading to
the concept of ``pion condensation'' in addition to the better known pion-loop
corrections, should we take the effective chiral quark theory to the next
level of sophistication. Finally, there are issues concerning the importance
of the instantons [12] when pions are involved.

First, we discuss briefly how Goldstone pions can be treated as ``quantized
fields''. This involves integration of the renormalization procedure as 
now widely adopted for chiral perturbation theory with the standard 
renormalization for quantized gauge field theories. This is not a trivial 
task which requires further studies. To a given order (in terms of, e.g., the
number of pion loops), it might be possible to obtain the ``renormalized'' 
effective chiral quark theory which can then be used for mapping onto 
the world of hadrons. For the purpose of the present paper, however, we
note that the problems which involve such renormalizations involve terms 
at least bi-linear in pion fields and thus are always of higher order 
in nature.

Next, we may discuss the changes when the constituent quark mass instead of the
very small current quark mass is adopted. As a qualitative estimate, we
approximate the first term of Eq. (24) by the first term of Eq. (16) $-$
we neglect a term of order $m^2 /p^2$ (with $p^2 \sim 1\, GeV^2$), a 
$(10 $-$ 20)$ \% effect. It affects most of the existing QCD sum rule 
predictions only in quantitative detail, but not qualitatively. This point
is also exemplified in detail in our analysis of the sum rules on the 
nucleon mass and the strong pion-nucleon coupling. 

It is clear that the proposal of matching the world of hadrons to that of the
effective chiral quark theory raises many interesting questions related to
fundamentals in QCD sum rules. It is also clear that it would be well beyond 
the scope of the present paper to address many such questions in detail. 
It is hoped that the encouraging results which we have obtained in the 
present paper would help to motivate further investigations along 
this direction.

To sum up, we have investigated, in the context of QCD sum rules, the 
role of Goldstone bosons arising from chiral symmetry breaking, 
especially on the need of matching the world of hadrons to that of some 
effective chiral quark theory.
We have shown how the proposal yields predictions on induced condensates 
which reproduce the observed value of the strong $\pi NN$ coupling 
constant. Our result indicate that the observed large $\pi NN$ coupling comes
primarily from the nonperturbative induced-condensate effect. On the other 
hand, the prediction on the parity-violating weak $\pi NN$ coupling,
$ f_{\pi NN} \sim (3.0\pm 0.5) \times 10^{-7}$, is
in good agreement with the value ($\sim 2 \times 10^{-7}$) obtained by 
Adelberger and Haxton [15] from anoverall fit to the existing data, but 
may appear to be slightly too large (but not yet of statistical 
significance) as compared to the most recent null experiment [18]. 
Altogether, we wish to argue that the proposal of matching onto a 
chirally broken effective theory should enable us to treat, in a 
quantitative and consistent manner, the various nonleptonic strong and 
weak processes involving Goldstone bosons.

\bigskip
\bigskip
\centerline{\bf Acknowledgement}
\medskip
\medskip

The author wishes to thank Ernest M. Henley, Xiangdong Ji, and Leonard S.
Kisslinger for very helpful conversations. He also wishes to
acknowledge Center for Theoretical Physics of MIT for
the hospitalities extended to him during his sabbatical leave. He also
acknowledges the National Science Council of R.O.C. for its partial support 
(NSC86-2112-M002-010Y) towards the present research.
 
\bigskip
\bigskip
 
\centerline{\bf References}

\item{1.} M. A. Shifman, A.J. Vainshtein, and V.I. Zakharov, Nucl. Phys. 
{\bf 147}, 385, 448 (1979).

\item{2.} T.-P. Cheng and L.-F. Li, {\it Gauge Theory of Elementary Particle
Physics} (Clarendon Press, Oxford, 1984); 
T.-Y. Wu and W-Y. P. Hwang, {\it Relativistic Quantum Mechanics and Quantum 
Fields} (World Scientific, Singapore, 1991).

\item{3.} B. L. Ioffe, Nucl. Phys. {\bf B188}, 317 (1981); (E) {\bf B191}, 591
(1981); V. M. Belyaev and B. L. Ioffe, Zh. Eksp. Teor. Fiz. {\bf 83}, 876 
(1982) [Sov. Phys. JETP {\bf 56}, 493 (1982)].

\item{4.} V. M. Belyaev and Ya. I. Kogan, Pis'ma Zh. Eksp. Teor. Fiz. {\bf 37},
611 (1983) [JETP Lett. {\bf 37}, 730 (1983]; C. B. Chiu, J. Pasupathy, and S.J.
Wilson, Phys. Rev. {\bf D32}, 1786 (1985); E. M. Henley, W-Y. P. Hwang, and
L.S. Kisslinger, Phys. Rev. {\bf D46}, 431 (1992); Chinese J. Phys. (Taipei)
{\bf 30}, 529 (1992).

\item{5.} S. Weinberg, Phys. Lett. {\bf B251}, 288 (1990); Physica {\bf 96A},
327 (1979).

\item{6.} H. Goergi, {\it Weak Interactions and Modern Particle Theory} 
(Benjamin/Cummings Publishing Co., Menlo Park, California, 1984);
A. Manohar and H. Georgi, Nucl. Phys. {\bf B234}, 189 (1984).

\item{7.} J. Gasser and H. Leutwyler, Phys. Rep. {\bf 87}, 77 (1982);
Nucl. Phys. {\bf B250}, 465 (1985). 

\item{8.} B.L. Ioffe, Nucl. Phys. {\bf B188}, 317 (1981); {\bf B191},
591 (E) (1981); Z. Phys. {\bf C18}, 67 (1983). 

\item{9.} E. M. Henley, W-Y. P. Hwang, and L. S. Kisslinger, Phys. Lett. {\bf
B} 367, 21 (1996).

\item{10.} K.-C. Yang, W-Y. P. Hwang, E.M. Henley, and L.S. Kisslinger, 
Phys. Rev. {\bf D47}, 3001 (1993).

\item{11.} S. Narison, Phys. Lett. {\bf B} 387, 162 (1996).

\item{12.} T. Sch\"afer and E.V. Shuryak, 1996 preprint hep-ph/9610451.

\item{13.} J.-L. Jacquot and J. Richert, Z. Phys. {\bf C} - Particles and
Fields, {\bf 56}, 201 (1992).

\item{14.} W-Y. P. Hwang, Ze-sen Yang, Y.S. Zhong, Z.N. Zhou, and
Shi-lin Zhu, 1996 preprint hep-ph/9610412.

\item{15.} See, e.g., E.G. Adelberger and W.C. Haxton, Ann. Rev. Nucl. Part.
Sci. {\bf 35}, 501 (1985); and J. Lang {\it et al.}, Phys. Rev. {\bf 34}, 1545
(1986).

\item{16.} B. Desplanques, J.F. Donoghue, and B.R. Holstein, Ann. Phys. (NY)
{\bf 124}, 449 (1980).

\item{17.} E.M. Henley, Ann. Rev. Nucl. Sci. {\bf 19}, 367 (1969; Chinese J.
Phys. {\bf 30}, 1 (1992).

\item{18.} C.A. Barnes {\it et al.} Phys. Rev. Lett. {\bf 40}, 840 (1978);
H.C. Evans {\it et al.}, Phys. Rev. Lett. {\bf 55}, 791 (1985); Phys. Rev.
{\bf C35}, 1119 (1987); M. Bini {\it et al.}, Phys. Rev. Lett. {\bf 55}, 795
(1985); Phys. Rev. {\bf C38}, 1195 (1988).

\item{19.} K. K. Gupta et al., Phys. Rev. {\bf D42}, 1604 (1990).

\bye